\documentclass[twocolumn,superscriptaddress,amsmath,amssymb,prl,longbibliography]{revtex4-1}

\usepackage{amssymb,amsmath}
\usepackage{cmap}
\usepackage{graphicx}
\usepackage{bm}
\usepackage{color}
\usepackage{blindtext}
\usepackage{hyperref}
\usepackage{listings}
\usepackage{booktabs}
\usepackage{multirow}
\usepackage{amsmath}
\usepackage{mathrsfs}
\usepackage{braket}
\usepackage{mathtools}

\usepackage{dsfont}
\usepackage{tabularx}

\begin{document}

\title{Elastic Spin and Orbital Angular Momenta}

\author{Konstantin Y. Bliokh}
\affiliation{Theoretical Quantum Physics Laboratory, RIKEN Cluster for Pioneering Research, Wako-shi, Saitama 351-0198, Japan}


\begin{abstract}
Motivated by recent theoretical and experimental interest in the spin and orbital angular momenta of elastic waves, we revisit canonical wave momentum, spin, and orbital angular momentum in isotropic elastic media. We show that these quantities are described by simple universal expressions, which differ from the results of [G. J. Chaplain {\it et al}., Phys. Rev. Lett. {\bf 128}, 064301 (2022)] and do not require separation of the longitudinal and transverse parts of the wavefield. For cylindrical elastic modes, the normalized $z$-component of the total (spin+orbital) angular momentum is quantized and equals the azimuthal quantum number of the mode, while the orbital and spin parts are not quantized due to the spin-orbit geometric-phase effects. In contrast to the claims of the above article, longitudinal, transverse, and `hybrid' contributions to the angular momenta are equally important and cannot be neglected. As another application of the general formalism, we calculate the transverse spin angular momentum of a surface Rayleigh wave. 
\end{abstract}


\maketitle

{\it Introduction.---}
Recently, there was a great renewed interest in the spin and orbital angular momenta of acoustic waves, i.e., sound waves in fluids or gases \cite{Shi2019,Toftul2019,Long2020,Burns2020,Long2020_II,Wei2020,Wang2021NC,Bliokh2022} and elastic waves in solids \cite{Zhang2014,Nakane2018,Long2018,Streib2021,Yuan2021,Sonner2021,Chaplain2022,Chaplain2022_II}. Although general field-theory principles of the momentum and angular momentum of sound and elastic waves have been known \cite{Jones1973,Soper,Lazar2007}, and the orbital angular momentum of sound waves was extensively studied both theoretically and experimentally \cite{Hefner1999,Thomas2003,Lekner2006,Volke2008,Zhang2011,Demore2012,Anhauser2012,Wunenburger2015, Wang2018,Rondon2019}, new observable phenomena involving spin-polarization properties of sound waves, as well as the spin and orbital angular momenta of elastic waves, prompted a new wave of investigations in these fields. 

This work is motivated by the recent papers \cite{Chaplain2022,Chaplain2022_II} which calculated the orbital angular momentum of elastic wave modes in an isotropic cylinder and reported its observation. We revisit the canonical momentum, spin, and orbital angular momentum of elastic waves in an isotropic medium using the field-theory approach \cite{Soper,Nakane2018,Burns2020,Bliokh2022}, which has been successfully employed in optics \cite{Berry2009,Bliokh2013NJP,Bliokh2014NC,Bliokh2015PR,Bliokh2017PRL,Bliokh2022_II}. We find that theoretical results of Refs.~\cite{Chaplain2022,Chaplain2022_II} contain significant errors, and show that the canonical momentum and angular momentum of elastic waves are described by rather simple general expressions, which do not require separation of the longitudinal (compression) and transverse (shear) parts of the wavefield. Moreover, the longitudinal, transverse, and `hybrid' contributions contained in these general expressions are equally important and cannot be neglected (as it was done in Refs.~\cite{Chaplain2022,Chaplain2022_II}). 

We apply general theory to the important case of elastic eigenmodes of a cylindrical waveguide, and find that their momentum and angular momentum properties are entirely analogous to those of optical guided modes \cite{Picardi2018}.
As another example, we calculate the transverse spin (previously explored in electromagnetic and sound evanescent waves \cite{Bliokh2014NC,Bliokh2015PR,Aiello2015,Bliokh2015NP,Lodahl2017,Bliokh2017PRL,Shi2019,Bliokh2019}) of a surface Rayleigh wave \cite{Long2018,Yuan2021,Sonner2021}.
Our results illuminate the universal character of the canonical momentum, spin, and orbital angular momentum of classical waves: not limited to transverse electromagnetic or longitudinal sound waves but equally applicable to mixed elastic waves.

It is worth remarking that the canonical momentum and angular momentum of acoustic waves are sometimes called `pseudomomentum' and `angular pseudomomentum', because these quantities are associated with translations and rotations of the wavefield with respect to the motionless medium \cite{Nakane2018,Streib2021,McIntyre1981,Peierls_I,Peierls_II,Stone2002,Gordon1973,Nelson1991}. 

{\it Canonical momentum, spin, and orbital angular momentum of elastic waves.---}
To start with, elastic waves in an isotropic medium can be described using the Lagrangian density \cite{Nakane2018,Chaplain2022} 
\begin{equation}
\label{eq_L}
{\cal L} = \dfrac{1}{2}\rho\, \dot{\bf a}^2 - U \equiv T - U\,. 
\end{equation}
%
where $\rho$ is the density of the medium, ${\bf a}({\bf r},t)$ is the displacement field, the dot stands for the time derivative (so that $\dot{\bf a} ={\bf v}$ is the velocity), $U$ is the potential energy density (involving spatial derivatives of ${\bf a}$ and {\it not} used explicitly in the calculations below), and $T$ is the kinetic energy density. 
 
The energy density $W$ and canonical momentum density ${\bf P}$ can be derived from the Lagrangian (\ref{eq_L}) by applying Noether's theorem with respect to space-time translations $t \to t + \delta t$ and ${\bf r} \to {\bf r} + \delta{\bf r}$ \cite{Soper,Bliokh2013NJP,Nakane2018,Bliokh2022_II}. This yields
\begin{eqnarray}
&&W= \frac{\partial {\cal L}}{\partial  \dot{\bf a}} \cdot  \dot{\bf a} -  {\cal L} = T + U \,, \nonumber \\
\label{eq_T}
&&{\bf P} = - \frac{\partial {\cal L}}{\partial  \dot{\bf a}} \cdot (\bm \nabla) {\bf a} = -\rho\, {\bf v}\cdot ({\bm \nabla}){\bf a}\,.
\end{eqnarray}
Similarly, the canonical angular momentum density ${\bf J}$ follows from the Lagrangian density (\ref{eq_L}) and Noether's theorem with respect to spatial rotations \cite{Soper,Bliokh2013NJP,Nakane2018,Bliokh2022_II}:
\begin{eqnarray}
\label{eq_J}
{\bf J} = - \frac{\partial {\cal L}}{\partial \dot{\bf a}} \cdot ({\bf r} \times {\bm \nabla}) {\bf a} -
\frac{\partial {\cal L}}{\partial \dot{\bf a}} \times {\bf a} \equiv {\bf L}  +{\bf S}\,, 
\end{eqnarray}
where the orbital and spin parts are
\begin{equation}
\label{eq_LS}
{\bf L} = {\bf r} \times {\bf P}\,,\quad  
{\bf S} = -\rho\, {\bf v}\times{\bf a}\,. 
\end{equation}

Equations (\ref{eq_L})--(\ref{eq_LS}) are in agreement with the results of \cite{Nakane2018,Streib2021} for the pseudomomentum and angular pseudomomentum of phonons. They also have forms entirely similar to the canonical momentum and angular momentum of sound waves in fluids or gases \cite{Jones1973,Shi2019,Burns2020,Bliokh2022}, but in contrast to the purely longitudinal sound waves (with ${\bm \nabla}\times {\bf a} = {\bf 0}$), elastic waves can have both longitudinal and transverse  contributions: ${\bf a}= {\bf a}_L + {\bf a}_T$, ${\bm \nabla}\times {\bf a}_L = {\bf 0}$, ${\bm \nabla}\cdot {\bf a}_T = 0$.

At the same time, Eqs.~(\ref{eq_T})--(\ref{eq_LS}) differ from the recent results of Refs.~\cite{Chaplain2022,Chaplain2022_II}.
To show this, from now on we consider monochromatic wavefields and introduce the complex-amplitude representation for the wavefields: ${\bf a}({\bf r},t) = {\rm Re}\!\left[ {\bm{\mathsf a}}({\bf r}) e^{-i\omega t} \right]$,
${\bf v}({\bf r},t) = {\rm Re}\!\left[ {\bm{\mathsf v}}({\bf r}) e^{-i\omega t} \right]$, where $\bm{\mathsf v} = -i\omega \bm{\mathsf a}$. Substituting this representation into Eqs.~(\ref{eq_T})--(\ref{eq_LS}) and performing time averaging over oscillations with frequency $\omega$, we obtain the time-averaged kinetic energy, momentum, and angular momentum densities:
\begin{eqnarray}
\bar{T} & = & \frac{\rho\,\omega^2}{4} |{\bm{\mathsf a}}|^2, \quad  
\bar{\bf P} = \frac{\rho\,\omega}{2}\, {\rm Im}\!\left[{\bm{\mathsf a}}^*\!\cdot ({\bm \nabla}){\bm{\mathsf a}}\right], \nonumber \\
\bar{\bf L} & = & {\bf r} \times \bar{\bf P}\,,\quad  
\bar{\bf S} = \frac{\rho\,\omega}{2} {\rm Im}\!\left({\bm{\mathsf a}}^*\!\times{\bm{\mathsf a}}\right). 
\label{eq_mono}
\end{eqnarray}
These are universal forms (also appearing for electromagnetic and sound waves \cite{Berry2009,Bliokh2013NJP,Bliokh2014NC,Bliokh2015PR,Bliokh2017PRL,Shi2019,Burns2020,Bliokh2022}) which resemble local expectation values of quantum-mechanical momentum ($-i{\bm \nabla}$), orbital angular momentum ($-i {\bf r} \times {\bm \nabla}$), and spin-1 ($-i\, \times$) operators with the `wavefunction' 
${\bm \psi} = \sqrt{\rho\omega/2}\, {\bf a}$, ${\bm \psi}^*\!\cdot {\bm \psi} = 2 \bar{T}$.

Equations (\ref{eq_mono}) are self-sufficient for an arbitrary elastic wavefield, including longitudinal and transverse parts. Substituting ${\bm{\mathsf a}} = {\bm{\mathsf a}}_L + {\bm{\mathsf a}}_T$ into Eqs.~(\ref{eq_mono}), one can see that all the quadratic quantities include longitudinal, transverse, and hybrid contributions \cite{Long2018,Chaplain2022}. For example, the canonical momentum density takes the form $\bar{\bf P} = \bar{\bf P}_L + \bar{\bf P}_T + \bar{\bf P}_H$:
\begin{eqnarray}
\bar{\bf P}_L & = & \frac{\rho\,\omega}{2}\, {\rm Im}\!\left[{\bm{\mathsf a}}_L^*\!\cdot ({\bm \nabla}){\bm{\mathsf a}}_L\right], \quad \bar{\bf P}_T = \frac{\rho\,\omega}{2}\, {\rm Im}\!\left[{\bm{\mathsf a}}_T^*\!\cdot ({\bm \nabla}){\bm{\mathsf a}}_T\right]
\nonumber \\
\bar{\bf P}_H & = & \frac{\rho\,\omega}{2}\, {\rm Im}\!\left[{\bm{\mathsf a}}_L^*\!\cdot ({\bm \nabla}){\bm{\mathsf a}}_T\right]
+ \frac{\rho\,\omega}{2}\, {\rm Im}\!\left[{\bm{\mathsf a}}_T^*\!\cdot ({\bm \nabla}){\bm{\mathsf a}}_L\right].
\label{eq_LTH}
\end{eqnarray}
These equations differ significantly from Eqs.~(12) in Ref.~\cite{Chaplain2022} and the corresponding orbital angular momentum calculated there. First, Eqs.~(12) have extra prefactors of squared speeds of the longitudinal and transverse waves, $c_p^2$ and $c_s^2$. These prefactors appeared due to a confusion between the momentum densities and the energy flux densities, and this resulted in the incorrect dimensionality of the main angular-momentum equation (15) in Ref.~\cite{Chaplain2022}. Second, Eqs.~(12) in Ref.~\cite{Chaplain2022} contain incorrect quadratic forms ${\rm Im}\!\left[({\bm{\mathsf a}}^*\!\cdot {\bm \nabla}){\bm{\mathsf a}}\right]$ instead of ${\rm Im}\!\left[{\bm{\mathsf a}}^*\!\cdot ({\bm \nabla}){\bm{\mathsf a}}\right]$. Since ${\bm{\mathsf a}}^*\!\cdot ({\bm \nabla}){\bm{\mathsf a}} = ({\bm{\mathsf a}}^*\!\cdot {\bm \nabla}){\bm{\mathsf a}} + {\bm{\mathsf a}}^*\!\times ({\bm \nabla} \times {\bm{\mathsf a}})$, these forms are equivalent only for the longitudinal contribution, but not for the transverse and hybrid ones. 

Next, Refs.~\cite{Chaplain2022,Chaplain2022_II} neglected the transverse and hybrid contributions to the orbital angular momentum of cylindrical eigenmodes. We argue that this unjustified for two reasons: (i) calculations were performed using erroneous general equations and (ii) it was assumed that the properly defined orbital angular momentum should be quantized and proportional to the azimuthal quantum number $\ell$. Below we show that this is not so: the azimuthal number $\ell$ determines the quantized {\it total} angular momentum (where the longitudinal, transverse, and hybrid contributions are equally important), while the properly defined orbital angular momentum necessarily contains $\ell$-independent part due to the spin-orbit coupling; this phenomenon is well known for optical and sound waves \cite{Bliokh2015NP,Bliokh2010,Picardi2018,Bliokh2019_II}.

{\it Momentum and angular momentum of cylindrical modes.---}
We now consider eigenmodes of an isotropic elastic cylinder \cite{Auld_book,Chaplain2022}. Notably, the calculations below are applicable to {\it any} cylindrically symmetric modes and universal quadratic forms (\ref{eq_mono}), independently of the nature of waves. The only property we use is that the cylindrical field has the form
\begin{equation}
{\bm{\mathsf a}} = \left[ a_r (r), a_\varphi (r), a_z (r) \right] e^{i\ell\varphi + i k_z z}\,,
\label{eq_mode}
\end{equation}
where $(r,\varphi,z)$ are the natural cylindrical coordinates, $k_z$ is the longitudinal wavenumber, and $\ell$ is the integer azimuthal quantum number. For the straightforward application of Eqs.~(\ref{eq_mono}), it is instructive to use the Cartesian field components $(a_x,a_y,a_z)$ and the associated basis of circular polarizations in the $(x,y)$-plane: $a^\pm = \dfrac{a_x \mp i a_y}{\sqrt{2}}=  \dfrac{a_r \mp i a_\varphi}{\sqrt{2}} e^{\mp i \varphi}$ \cite{Picardi2018}. Here the geometric-phase factors $e^{\mp i \varphi}$ originate from the rotation of the cylindrical coordinates with respect to the Cartesian ones \cite{Bliokh2015NP}. In the circular basis, the operator of the $z$-component of the spin becomes diagonal: ${\rm diag}(1,-1,0)$.	

Substituting the field (\ref{eq_mode}) in the circular-Cartesian basis into Eqs.~(\ref{eq_mono}), we obtain the normalized $z$-components of the momentum, spin, and orbital angular momentum densities:
\begin{eqnarray}
\frac{\bar{P}_z}{2\bar{T}} & = & \frac{k_z}{\omega}\,,\quad\quad \frac{\bar{J}_z}{2\bar{T}} = \frac{\ell}{\omega}\,, \nonumber \\
\frac{\bar{L}_z}{2\bar{T}} & = & \frac{\ell}{\omega} - \frac{|a^+|^2 - |a^-|^2}{\omega |{\bm{\mathsf a}}|^2}\,,\quad
\frac{\bar{S}_z}{2\bar{T}} = \frac{|a^+|^2 - |a^-|^2}{\omega |{\bm{\mathsf a}}|^2}\,.
\label{eq_mode_PLS}
\end{eqnarray}
%
Equations (\ref{eq_mode_PLS}) are entirely similar to their optical counterparts derived in Ref.~\cite{Picardi2018} (apart from the normalization to the double-kinetic energy density $2\bar{T}$ instead of the total energy density $\bar{W}$), which reflects their universal character. These equations show that the total (spin+orbital) angular momentum is quantized, and $\ell$ is the {\it total} rather than orbital angular momentum quantum number \cite{Picardi2018}. The division of the total angular momentum into the orbital and spin parts is closely related to the {\it geometric and dynamical phases} calculated for the vector field ${\bf a}$ along circular contours $r={\rm const}$. This is explained in detail in Refs.~\cite{Picardi2018,BAD2019}, so we do not repeat these considerations here and only note that the $\ell$-independent spin-dependent second term in the orbital angular momentum $\bar{L}_z$ in Eqs.~(\ref{eq_mode_PLS}) can be regarded as a manifestation of the {\it spin-orbit interaction} \cite{Bliokh2015NP,Bliokh2010}. 

We emphasize that Eqs.~(\ref{eq_mode_PLS}) have simple meaningful forms in terms of the total elastic wavefield ${\bm{\mathsf a}}$, including both longitudinal and transverse parts, and it does not make much sense to separate the longitudinal, transverse, and hybrid contributions there. These are all equally important for the resulting momentum and angular momentum densities. Furthermore, although longitudinal and transverse elastic waves in a bulk solid propagate with different velocities $c_p$ and $c_s$, the cylindrical guided mode is a {\it single mixed mode} which propagates with the phase velocity $v_{ph} = \omega / k_z$, and its normalized linear momentum has the corresponding well-defined value $|\bar{P}_z|/(2\bar{T}) = 1/v_{ph}$. This also evidences that separation of the longitudinal and transverse parts of the field has an artificial technical character in this case. In fact, no local measurement can distinguish between the longitudinal, transverse, or hybrid contributions to the momentum or angular momentum densities in the field: any probe interacts with the total local field ${\bm{\mathsf a}}({\bf r})$, {where each Cartesian component generally contains contributions from ${\bf a}_L$ and ${\bf a}_T$}.

It is worth noticing that the normalized spin density in Eqs.~(\ref{eq_mode_PLS}) is limited as $\omega |\bar{S}_z|/(2\bar{T}) \leq 1$, whereas the orbital and total angular momenta are only limited by the value of $\ell$: $\omega |\bar{L}_z|/(2\bar{T}) \leq |\ell |$, $\omega |\bar{J}_z|/(2\bar{T)} \leq |\ell |$. Note also that we calculated the spatial densities of all quantities; their integral values can be obtained via the integration over the transverse cross section: $\langle \bar{L}_z \rangle = \int \bar{L}_z\, dx dy$, etc. Notably, the integral value of the total energy can be written as $\langle \bar{W} \rangle = 2 \langle \bar{T} \rangle$ (see \cite{Auld_book}, Section 10-O), so that the integral versions of Eqs.~(\ref{eq_mode_PLS}) involve the integral momentum and angular momenta normalized by $\langle \bar{W} \rangle$.

{\it Transverse spin of a Rayleigh wave.---}
Let us consider another application of general Eqs.~(\ref{eq_mono}) to an inhomogeneous elastic wave: the {\it transverse spin} of a surface Rayleigh wave \cite{Long2018,Sonner2021}. The transverse spin of surface or evanescent waves is an interesting phenomenon which recently attracted great attention in optics and acoustics \cite{Bliokh2014NC,Bliokh2015PR,Aiello2015,Bliokh2015NP,Lodahl2017,Bliokh2017PRL,Long2018,Shi2019, Bliokh2019,Yuan2021,Sonner2021}.  
The field of the Rayleigh wave propagating along the $z$-axis along the $x=0$ surface of an isotropic medium ($x<0$) can be written as \cite{LL_elasticity} ${\bm{\mathsf a}} = (a_x,0,a_z)$,
\begin{eqnarray}
a_x & \propto & i \left(\alpha k_z e^{\kappa_t x} -\kappa_l  e^{\kappa_l x} \right) e^{ik_z z} \equiv i A(x) e^{ik_z z} \,, \nonumber \\
a_z & \propto & \left(k_z e^{\kappa_l x} - \alpha \kappa_t  e^{\kappa_t x} \right) e^{ik_z z} \equiv B(x) e^{ik_z z} \,,
\label{eq_Rayleigh}
\end{eqnarray}
where $\kappa_{l,t} = \sqrt{k_z^2 - \omega^2/c_{l,t}^2}$, $\alpha = \dfrac{2-\xi^2}{2\sqrt{1-\xi^2}}\, {\rm sgn}(k_z)$, and $\xi = \dfrac{\omega}{kc_t}$. Substituting the field (\ref{eq_Rayleigh}) into Eqs.~(\ref{eq_mono}), 
we readily find that the Rayleigh wave carries longitudinal momentum and transverse $y$-directed spin angular momentum:
\begin{eqnarray}
\frac{\bar{P}_z}{2\bar{T}} = \frac{k_z}{\omega}\,,\quad
\frac{\bar{S}_y}{2\bar{T}} = \frac{2 AB}{\omega(A^2 + B^2)}\,,
\label{eq_Rayleigh_PS}
\end{eqnarray}
Akin to the optical and sound-wave transverse spin, the elastic transverse spin (\ref{eq_Rayleigh}) and (\ref{eq_Rayleigh_PS}) flips its sign with the flip of $k_z$. This `spin-momentum locking' \cite{Bliokh2015Science,Mechelen2016} is used for the efficient spin-direction coupling \cite{Bliokh2015PR,Aiello2015,Bliokh2015NP,Lodahl2017,Shi2019,Long2018,Yuan2021,Rodrguez2013,Petersen2014,Feber2015}.
As in the previous example of a cylindrical waveguide, no local measurement of the spin or polarization can distinguish between longitudinal, transverse, or hybrid contributions. Any probe interacts with the total elliptically-polarized displacement field ${\bm{\mathsf a}}({\bf r})$, Eq.~(\ref{eq_Rayleigh}), where the longitudinal and transverse parts both have qualitively-similar elliptical polarizations.

{\it Conclusions.---}
Thus, we have revisited the canonical momentum, spin, and orbital angular momentum of elastic waves in an isotropic medium. Using rather general approach we were able to derive these quantities without explicit use of the potential elastic energy, equations of motion, Lam\'{e} coefficients, etc. In particular, we have showed that the canonical momentum and angular momentum densities have universal forms independent of either transverse, or longitudinal, or mixed character of the wavefield. We have also calculated the normalized $z$-components of the momentum, spin, and orbital angular momentum of elastic eigenmodes of an isotropic cylinder. These are described by universal equations valid  for cylindrically-symmetric wavefields independently of their nature. As another example, we calculated the transverse spin of a surface Rayleigh wave. In both cases, the wavefield represents a single mixed mode, and it does not make much sense to separate the longitudinal, transverse, and hybrid contributions to its momentum and angular momentum. Our work provides significant corrections to the results of recent paper \cite{Chaplain2022}, as well as a theoretical basis for further studies of the momentum and angular momentum of elastic waves in isotropic media. 



\bibliography{References_elastic}

\end{document}